\documentclass[sigconf,table]{acmart}

\renewcommand\footnotetextcopyrightpermission[1]{}

\begin{document}
\fancyhead{}

\title{Recommenders with a mission: assessing diversity in news recommendations}



\author{Sanne Vrijenhoek}
\affiliation{Institute for Information Law, University of Amsterdam}
\email{s.vrijenhoek@uva.nl}
 
\author{Mesut Kaya}
\affiliation{Independent Researcher$^1$}
\email{mesutt.kayaa@gmail.com}
\thanks{$^1$This work was done while the author was at TU Delft.}
 
\author{Nadia Metoui}
\affiliation{Department of Communication Science, University of Amsterdam}
\email{n.metoui@uva.nl}
 
\author{Judith M{\"o}ller}
\affiliation{Department of Communication Science, University of Amsterdam}
\email{j.e.moller1@uva.nl}
 
\author{Daan Odijk}
\affiliation{RTL Nederland B.V.}
\email{daan.odijk@rtl.nl}
 
\author{Natali Helberger}
\affiliation{Institute for Information Law, University of Amsterdam}
\email{n.helberger@uva.nl}


\begin{abstract}
News recommenders help users to find relevant online content and have the potential to fulfill a crucial role in a democratic society, directing the scarce attention of citizens towards the information that is most important to them. Simultaneously, recent concerns about so-called filter bubbles, misinformation and selective exposure are symptomatic of the disruptive potential of these digital news recommenders. Recommender systems can make or break filter bubbles, and as such can be instrumental in creating either a more closed or a more open internet. Current approaches to evaluating recommender systems are often focused on measuring an increase in user clicks and short-term engagement, rather than measuring the user's longer term interest in diverse and important information.

This paper aims to bridge the gap between normative notions of diversity, rooted in democratic theory, and quantitative metrics necessary for evaluating the recommender system. We propose a set of metrics grounded in social science interpretations of diversity and suggest ways for practical implementations. 

\end{abstract}


%

\maketitle

\section{Introduction}
News recommender algorithms have the potential to fulfill a crucial role in democratic society. By filtering and sorting information and news, recommenders can help users to overcome maybe the greatest challenge of the online information environment: finding and selecting relevant online content - content they need to be informed citizens, be on top of relevant developments, and  have their say~\cite{eskens2017}. Informed by data on what a user likes to read, what people similar to him or her like to read, what content sells best, etc., recommenders use machine learning and AI techniques to make ever smarter suggestions to their  users~\cite{thurman2012,dorr2016,lewis2015,kunaver2017}.
For the news media, algorithmic recommendations offer a way to remain relevant on the global competition for attention, create higher levels of engagement with content, develop ways of informing citizens and offer services that people are actually willing to pay for~\cite{bodo2019}.
With this comes the power to channel attention and shape individual reading agendas and thus new risks and responsibilities. Recommender systems can be pivotal in deciding what kind of news the public does and does not see. Depending on their design, recommenders can either unlock the diversity of online information ~\cite{nguyen2014, helberger2019democratic_rs} for their users,
or lock them into boring routines of "more of the same", or in the most extreme case into so-called filter bubbles~\cite{pariser2011filter}
and information sphericules. 

The most frequently used key performance indicators, or KPIs, for optimizing recommender systems, assess and aim to maximize short-term user engagement, such as click-through rate or time spent on a page~\cite{jannach2019measuring}. Often, these KPIs are defined by data limitations, and by technological and business demands rather than the societal and democratic mission of the media. 
More recently however a process of re-thinking algorithmic recommender design has begun, in response to concerns from users~\cite{thurman2019}, regulators (e.g., EU HLEG~\cite{hleg2018ai}), academics, and news organizations themselves~\cite{bodo2019}. 
Finding ways to develop new metrics and models of more "diverse" recommendations has developed into a vibrant field of experimentation - in academia as well as in the data science and R\&D departments of a growing number of media corporations.

But what exactly does diverse mean, and how much diversity is 'enough'? As central as diversity (or pluralism, a notion that is often used interchangeably) is to many debates about the optimal design of news recommenders, as unclear it is what diverse recommender design actually entails. Designing for diversity in recommendation systems can be achieved with a value by design approach. Value by design has been defined as "a theoretically grounded approach to the design of technology that accounts for human values in a principled and comprehensive manner throughout the design process," or what Friedman et.al. called a "pro-active design component"~\cite{friedmansurvey2017}. A pro-active design component means finding ways to conceptualize and implement a value (here diversity) in a way that can inform the design of recommendation algorithms. This is distinct from research that seeks to measure (empirically) the effects of news recommenders on the diversity of contents that users are exposed to. Translating a value into concrete design requirements is not a straightforward process. It can involve going through several iterations of testing and adjusting the parameters or models that inform an algorithm. 
In the growing literature that tries to conceptualise and translate diversity into specific design requirements, a gap between the computer science and the normative literature can be observed. While diversity in the computer science literature is often defined as concrete technical metrics, such as the intra-list distance of recommended items~\cite{vargas2011rank, castells2015novelty} (see also Section~\ref{sec:background_tech}), diversity in the normative sense is about grand concepts: democracy, freedom of expressions, cultural inclusion, mutual respect and tolerance~\cite{helberger2019democratic_rs}. There is a mismatch between different theoretical understandings of the construct of diversity, similar to the one observed in Fairness research~\cite{jacobs2019measurement}.
For news recommenders to be truly able to unlock the abundance of information online and inform citizens better, it is imperative to find ways to overcome the fundamental differences in approaching diversity. There is a need to reconceptualise this central but also elusive concept in a way that both does justice to the goals and values that diversity must promote, as well as facilitates the translation of  diversity into metrics that are concrete enough to inform algorithmic design. 

This paper describes the efforts of a team from computer science, communication science, and media law and policy, to bridge this gap between normative and computational approaches towards diversity, and translate diversity, as a normative concept, to a concrete set of metrics that can be used to evaluate and/or compare different news recommender designs. 



We first survey diversity from a technical point of view (Section \ref{sec:background_tech}) and from a social science interpretation, including its role in democratic models (Section \ref{sec:background}). In Section \ref{sec:evaluation_metrics} we expand upon the social science notion of diversity, and propose five metrics grounded in Information Retrieval that reflect our normative approach. We cover the limitations of the proposed metrics and this approach in Section \ref{sec:limitations}. We conclude with detailing our implementation of the metrics and the steps to undertake as a media company when intending to adopt this normative notion of diversity in practice.
\label{sec:introduction}

\section{A technical conception of diversity in news recommenders}
\label{sec:background_tech}
Typically, generating a recommendation is seen as a reranking problem. Given a set of candidate items, the goal is to present these items in such a way that the user finds the item he or she is most interested in at the top, followed by the second-most interesting one, and so on. How well this recommendation reflects the actual interest of the user is called the \emph{accuracy} of the recommendation. \emph{Content-based} approaches aim to maximize this accuracy by looking at the type of items that the user has interacted with before and recommend similar ones. In the context of news recommendations, one could think of finding topics or overall texts that are similar to what is in the user’s reading history. On the other hand, in \emph{collaborative filtering} approaches, the algorithm considers what other users similar to the user in question have liked, and recommends those. Most state-of-the-art systems are hybrids of these approaches. Evaluation of the system can be done in both an online and offline fashion; offline often includes testing the system on a piece of held-out data on its accuracy, whereas online evaluation monitors for increases or decreases of user interactions and click-through rates following the issued recommendations~\cite{10.1145/2532508.2532511}.\\
However, this approach by its definition unduly promotes the items similar to what a user has seen before, locking the user in a feedback loop of "more of the same"~\cite{mcnee2006being}. It also introduces a so-called "confounding bias"~\cite{10.1145/3240323.3240370}, which happens when an algorithm attempts to model user behavior when the algorithm itself influences that behavior. 
To tackle this in many currently operational systems "beyond-accuracy" metrics diversity, novelty, serendipity and coverage are introduced.
\emph{Diversity} reflects how different the items within the recommendation set are from each other. One intuitive usecase can be found in the context of ambiguous search queries. A user searching for "orange" should receive results about the color, the fruit, the telecom company, the Dutch royal family, and the river in Namibia, and not just about the one the system thinks he or she is most likely to be interested in. The challenge then lies in how to define this difference or distance. In the context of news recommendations many different approaches exist, such as using a cosine similarity on a bag of words model or by calculating the distance between the article's topics~\cite{ziegler2005improving}. \\
The concepts of \emph{novelty} and \emph{serendipity} are strongly linked. \emph{Novelty} reflects the likeliness that the user has never seen this item before, whereas \emph{serendipity} reflects whether a user was positively surprised by the item in question. However, an item can be novel without being serendipitous (such as the weather forecast), and an item may also be serendipitous without being novel (such as an item that has been seen a long time ago, but becomes relevant again in light of recent events). A common approach to improving novelty and serendipity is by unlocking the "long tail" content of the  system, while still optimizing for user accuracy. The long tail refers to the "lesser known" content in the system, that is less popular and therefore seen by less users. By recommending less popular content the recommender systems increase the chance that an item is actually novel to a user. \\
Lastly, \emph{coverage} reflects to what extent all the items available in the system have been recommended to at least a certain number of users. This metric is naturally strongly influenced by the novelty of the recommendations, as increasing the visibility of lesser-seen items increases the overall coverage of all items.

\section{A democrative conception of diversity in news recommenders}
\label{sec:background}
What becomes apparent from the overview in Section \ref{sec:background_tech} is that although there are various attempts to conceptualize evaluation metrics beyond accuracy in the computer science literature, these metrics are constructed for the broad field of recommendation systems, and are therefore not only relevant in the context of news, but also for music, movies, web search queries and even online dating. However, what they win in generalizability, they lose in specificity. They are not grounded in, and do not refer back to the normative understanding of diversity in the media law, fundamental rights law, democratic theory and media studies/communication science literature.

Before we define more quantitative metrics to assess diversity in news recommendation, we first offer a conceptualization of diversity. 
Following the definition of the Council of Europe, diversity is not a goal in itself, it is a concept with a mission, and it has a pivotal role in promoting the values that define us as a democratic society. These values may differ according to different democratic approaches.
This article builds on a conceptualisation of diversity in recommendations that have been developed by~\citet{helberger2019democratic_rs}. Here, ~\citeauthor{helberger2019democratic_rs} combines the normative understanding of diversity, meaning what should diverse recommendations look like, with more empirical conceptions, meaning what is the impact of diverse exposure on users.
There are many theories of democracy, but the paper by ~\citeauthor{helberger2019democratic_rs} focuses on 4 of the most commonly used theories when talking about the democratic role of the media: \emph{Liberal}, \emph{Participatory}, \emph{Deliberative} and \emph{Critical} theories of democracy (see also~\cite{christians2009,dahlberg2011,karppinen2011conceptions,stromback2005}).

It is important to note that no model is inherently better or worse than another. Which model is followed is something that should be decided by the media companies themselves, following their mission and dependent on the role they want to play in a democratic society.

\subsection{The Liberal model}
\label{sec:liberal}
In liberal democratic theory, individual freedom, including fundamental rights such as the right to privacy and freedom of expression, dispersion of power but also personal development and autonomy of citizens stands central. The liberal model is in principal sympathetic to the idea of algorithmic recommendations and considers recommenders as tools to enable citizens to further their autonomy and find relevant content. 
The underlying premise is that citizens know for themselves best what they need in terms of self-fulfillment and exercising their fundamental rights to freedom of expression and freedom to hold opinions, and even if they do not, this is only to a limited extent a problem for democracy. This is because the normative expectations of what it means to be a good citizen are comparatively low and there is a strict division of tasks, in which \emph{"political elites [...] act, whereas citizens react"}\cite{stromback2005}. 
 
Under such liberal perspective, diversity would entail a user-driven approach to diversity that reflects citizens interests and preferences not only in terms of content, but also in terms of for example style, language and complexity.
The \emph{liberal recommender} is required to inform citizens about prominent issues, especially during key democratic moments such as election time, but else it is expected to take little distance from personal preferences. It is perfectly acceptable for citizens to be consuming primarily cat videos and celebrity news, as long as doing so is an expression of their autonomy.

\subsubsection*{Summary.} The liberal model of democracy promotes self-development and autonomous decision making. As such, a news recommender following a liberal approach should focus on the following criteria:
\begin{itemize}
 \item Facilitating the specialization of a user in an area of his/her choosing
 \item Tailored to a user’s preferences, both in terms of content and in terms of style
\end{itemize}

\subsection{The Participatory model}
An important difference between the liberal and the participatory model of democracy is what it means to be a good citizen. Under participatory conceptions, the role of (personal) freedom and autonomy is to further the common good, rather than personal self-development~\cite{held2006models}.
Citizens cannot afford to be uninterested in politics because they have an active role to play in helping the community to thrive~\cite{stromback2005}.
Accordingly, the media, and by extension news recommenders must do more than to give citizens 'what they want', and instead provide citizens with the information they need to play their role as active and engaged citizens~\cite{baker1998,ferre2018,karppinen2013,tandoc2015}, and to further the participatory values, such as inclusiveness, equality, participation, tolerance.
\emph{Participatory recommenders} must also pro-actively address the fear of missing out on important information and depth, and the concerns about being left out. Here the challenge is to make a selection that gives a fair representation of different ideas and opinions in society, while also helping a user to gain a deeper understanding, and feeling engaged, rather than confused.
This also involves that recommenders are able to respond to the different needs of users in which information is being presented. The form of presentation is an aspect that is often neglected in discussions around news recommender diversity, ignoring the fact that different people have different preferences and cognitive abilities to process information. Accordingly, the media should 'frame politics in a way that mobilizes people's interests and participation in politics'. \citet{stromback2005} and \citet{ferree2002four} speak of 'empowerment': to be truly empowering, media content needs to be presented in different forms and styles \cite{ferree2002four, christians2006media, zaller2003new}. By extension, this means that diversity is not only a matter of the diversity of content, but also of communicative styles.
What would then characterize diversity in a participatory recommender are, on the one hand, active editorial curation in the form of drawing attention to items that citizens 'should know', taking into account inclusive and proportional representation of main political/ideological viewpoints in society; a focus on political content/news, but also: non-news content that speaks to broader public and, on the other hand, a heterogeneity of styles and tones, possibly also emotional, empathetic, galvanizing, reconciliatory.

\subsubsection*{Summary.} The participatory model of democracy aims to enable people to play an active role in society. It values the idea of the ‘common good’ over that of the individual. Therefore, a participatory recommender should follow the following principles:
    \begin{itemize}
    \item Different users do not necessarily see the same articles, but they do see the same topics.
    \item Article’s complexity is tailored to a user’s preference and capability
    \item Reflects the prevalent voices in society
    \item Empathetic writing style 
    \end{itemize}
\subsection{The Deliberative model}
The participatory and the deliberative models of democracy have much in common (compare \citet{ferree2002four}). Also in the deliberative or discursive conceptions of democracy, community and active participation of virtuous citizens stands central. One of the major differences is that the deliberative model operates on the premise that ideas and preferences are not a given, but that instead we must focus more on the process of identifying and negotiating and, ultimately, agreeing on different values and issues \cite{ferree2002four, karppinen2011conceptions}.
Political and public will formation is not simply the result of who has the most votes or 'buyers', but it is the result of a process of public scrutiny and intensive reflection \cite{held2006models}. This involves a process of actively comparing and engaging with other also contrary and opposing ideas ~\cite{manin1987legitimacy}.
The epistemological shift from information to deliberation has important implications for the way the role of news recommenders can be conceptualised. Under a deliberative perspective, it is not enough to 'simply' inform people. The media need to do more, and has an important role in "promoting and indeed improving the quality of public life - and not merely reporting on and complaining about it"~\cite{christians2009}. \citet{stromback2005} goes even further and demands that the media should also "actively foster political discussions that are characterised by impartiality, rationality, intellectual honesty and equality among the participants". 
Diversity in the deliberative conception has the important task of confronting the audience with different and challenging viewpoints that they did not consider before, or not in this way~\cite{manin1987legitimacy}. 
Concretely, this means that a deliberative recommender (or recommendation) should include a higher share of articles presenting various perspectives, diversity of emotions, range of different sources; it should strive for equal representation, as well as on recommending items of balanced content, commentary, discussion formats, background information; potentially some prominence for public service media content (as the mission of many public service media includes the creation of a deliberative public sphere), as well as a preference for rational tone, consensus seeking, inviting commentary and reflection. 

\subsubsection*{Summary.}
The focus of the deliberative recommender is on presenting different opinions and values in society, with the goal of coming to a common consensus or agreeing on different values. 
\begin{itemize}
\item Focus on topics that are currently at the center of public debate
\item Within those topics, present a plurality of voices and opinions
\item Impartial and rational writing style
\end{itemize}

\subsection{The Critical model}
A main thrust of criticism of the deliberative model is that it is too much focused on rational choice, on drawing an artificial line between public and private, on overvaluing agreement and disregarding the importance of conflict and disagreement as a form of democratic exercise \cite{karppinen2013}. The focus on reason and tolerance muffles away the stark, sometimes shrill contrasts and hidden inequalities that are present in society, or even discourage them from developing their identity in the first place. Accordingly, under more radical or critical perspectives, citizens should look beyond the paint of civil and rational deliberation. They should discover and experience the many marginalised voices of those "who are 'outsiders within' the system"\cite{ferree2002four}, and when doing so critically reflect on reigning elites and their ability to give these voices their rightful place in society. Diverse critical recommenders hence do not simply give  people what they want. Instead, they actively nudge readers to experience otherness, and draw attention to the marginalised, invisible or less powerful ideas and opinions in society. And again, it is not only the question of what kinds of content are presented but also the how: whereas in the deliberative and also the participatory model, much focus is on a rational, reconciliary  and measured tone,  critical recommenders would also offer room for alternative forms of presentations: narratives that appeal to the 'normal' citizen because they tell an everyday life story, emotional and provocative content, even figurative and shrill tones  -  all with the objective to escape the standard of civility and the language of the stereotypical "middle-aged, educated, blank white man"\cite{young1996communication}.

\subsubsection*{Summary.} The critical recommender aims to provide a platform to those voices and opinions that would otherwise go unheard. From a critical democracy perspective on diversity, recommenders should be optimized on the following principles:
\begin{itemize}
\item Emphasis on voices from marginalized groups
\item Emotional writing style
\end{itemize}

\section{Diversity Metrics} 
\label{sec:evaluation_metrics}
The democratic models described in Section~\ref{sec:background} lead to different conceptualizations of diversity as a value, which again translate into different diversity expectations for recommender systems. In this section, we propose five metrics that follow directly from these expectations, grounded in democratic theory and adapted from existing Information Retrieval metrics:
\emph{Calibration}, \emph{Fragmentation}, \emph{Activation}, \emph{Representation} and \emph{Alternative Voices}. For each of these metrics, we explain the concept and link to democratic theory. Furthermore we make a suggestion for operationalization, but note that this work is an initial outline and that much work still needs to be done. Future work should include more work on the validity of the metrics, for example by following the measurement models specified in ~\citet{jacobs2019measurement}. Lastly we mention a number of the limitations of the currently proposed metrics and their operationalizations. 

Table \ref{tab:table_overview} provides an overview of the different models, metrics and their expected value ranges. Note that not all metrics are relevant to all models.

Before explaining the metrics, we define the following variables that are relevant to multiple metrics:
\begin{itemize}
    \item $p$: The list of articles the recommender system could make its selection from, also referred to as the 'pool'
    \item $q$: The unordered list of articles in the recommendation set
    \item $Q$: The ordered list of articles in the recommendation set
    \item $r$: The list of articles in a user's reading history
\end{itemize}

\subsection{Calibration}
\label{m:calibration}
The \emph{Calibration} metric reflects to what extent the issued recommendations reflect the user's preferences. A score of 0 indicates a perfect Calibration, whereas a higher score indicates a larger divergence from the user's preferences.
\subsubsection{Explanation.} Calibration is a well-known metric in traditional recommender system literature~\cite{steck2018}. It is calculated by measuring the difference in distributions of categorical information, such as topics in the news domain or genres in the movie domain, between what is currently recommended to the user and what the user has consumed in the past. However, we extend our notion of calibration beyond topicality or genre. News recommendations can also be tailored to the user in terms of article style and complexity, allowing the reader to receive content that is attuned to their information needs and processing preferences. This may be split up within different topics; a user may be an expert in the field of politics but less so in the field of medicine, and may want to receive more complex articles in case of the first, and less in case of the second.
\subsubsection{In the context of democratic recommenders.}
The Calibration metric is most significant for recommenders following the Liberal and Participatory model. The aim of the Liberal model is to facilitate user specialization, and assumes that the user eventually knows best what they want to read. In these models, we expect the Calibration scores to be closer to 0. On the other hand, the Participatory model favors the common good over the individual. We therefore expect a higher degree of divergence in Calibration, at least when considered in light of topicality. Both models, but especially the Participatory model, require that the user receives content that is tailored to their needs in terms of article complexity, and in this context we expect a Calibration score that is closer to zero.
\subsubsection{Operationalization.}
For the operationalization of a recommender's Calibration score it is important to have information on not only an article's topic and complexity, which can potentially be automatically extracted from an article's body (see for example \citet{feng2010comparison} and \citet{kim2011topic}), but also on the user's preferences regarding this matter. Note that topicality can be both generic (politics, entertainment, sports, etc) and more specific (climate change, Arsenal). In light of democratic theory more fine-grained information is preferable, but this is not always available. 
\citet{steck2018} uses the Kullback-Leibler divergence between two probability distributions as Calibration metric, as follows:
\vspace{-.25\baselineskip}
\[ Calibration_{(r,q)} = \sum_{c}^{} r(c|u) log \frac{r(c|u)}{\tilde{q}(c|u)} \]
where $r(c|u)$ is the distribution of categorical information $c$ across the  articles consumed by the user in the past, and $\tilde{q}(c|u)$ is an approximation of $q(c|u)$ (necessary since KL divergence diverges if $q(c|u)=0$), which is the  distribution of the categories c across the current recommendation set.
As mentioned before, a score of 0 indicates that there is no divergence between the two distributions, meaning they are identical. The higher the Calibration score, the larger the divergence. As KL divergence can yield very high scores when dividing by numbers close to zero, outliers can greatly influence the average outcome. Therefore, the aggregate Calibration score is calculated by taking the median of all the Calibration scores for individual users. 
\subsubsection{Limitations.} This approach is tailored to categorical data, but sometimes our data may be numerical rather than categorical, for example in the case of article complexity. In these cases, a simple distance measure may suffice over the more complex Kullback-Leibler divergence. 

\subsection{Fragmentation}
\label{m:fragmentation}
The \emph{Fragmentation} metric denotes the amount of overlap between news story chains shown to different users. A Fragmentation score of 0 indicates a perfect overlap between users, whereas a score of 1 indicates no overlap at all.
\subsubsection{Explanation.} News recommender systems create a recommendation by filtering from a large pool of available news items. By doing so they may stimulate a common public sphere, or create smaller and more specialized 'bubbles'. This may occur both in terms of topics recommended, which is the focus of the Fragmentation metric, and in terms of presented perspectives, which will be later explained in the Representation metric. Fragmentation specifically compares differences in recommended news story chains, or sets of articles describing the same issue or event from different perspectives, writing styles or points in time~\cite{nicholls2019}, \emph{between} users; the smaller the difference, the more aware the users are of the same events and issues in society, and the more we can speak of a joint agenda. When the news story chains shown to the users differ significantly, the public sphere becomes more fragmented, hence the term Fragmentation. 
\subsubsection{In the context of democratic recommenders.}
Both the Participatory and Deliberative models favor a common public sphere, and therefore a Fragmentation score that is closer to zero. The Liberal model on the other hand promotes the specialization of the user in their area of interest, which in turn causes a higher Fragmentation score. Finally the Critical model, with its emphasis on drawing attention to power imbalances prevalent in society as a whole, calls for a low Fragmentation score. 
\subsubsection{Operationalization.}
This metric requires that individual articles can be aggregated into higher-level news story chains over time. This can be done through manual annotation or automated extraction process. One unsupervised learning approach for doing this automatically can be found in~\citet{nicholls2019}. Here the articles are clustered in a similarity network based on tfidf or BM25F distance between articles, with a moving three-day window for scalability.
Once the stories are identified, the Fragmentation score can be defined as the aggregate average distance between all sets of recommendations between all users.
\citet{dillahunt2015detecting}, which aimed to detect filter bubbles in search engine results, defines this distance with the Kendall Tau Rank Distance (KDT), which measures the number of pairwise disagreements between two lists of ranked items. However, Kendall Tau is not suitable when the two lists can be (largely) disjointed. It also penalizes differences at the top of the list equally to those more at the bottom. Instead we base our approach on the Rank Biased Overlap used in \citet{webber2010similarity}:
\vspace{-.25\baselineskip}
\[ RBO(Q_{1},Q_{2},s)=(1-s)\sum_{d=1}^{\infty} s^{d-1}\cdot A_{d} \]
where $Q_{1}$ and $Q_{2}$ denote two (potentially) infinite ordered lists, or two recommendations issued to users 1 and 2, and $s$ a parameter that generates a set of weights with a geometric progression starting at 1 and moving towards 0 that ensures the tail of the recommendation is counted less severely compared to its head. Because of this there is a natural cut-off point where the score stabilizes. We iterate over the ranks $d$ in the recommendation set, and at each rank we calculate the average overlap $A_{d}$. 
Because Rank-Biased Overlap yields a score between 0 and 1, with 0 indicating two completely disjoint lists and 1 a perfect overlap, and the score that is expressed is semantically opposite of what we aim to express with the Fragmentation metric, we obtain the Fragmentation score by calculating 1 minus the Rank-Biased Overlap.
Lastly, the aggregate Fragmentation score is calculated by averaging the Fragmentation score between each user and every other user.
\subsubsection{Limitations.} Since this approach is computationally expensive (every user is compared to every other user, which is $O(n^{2})$ complexity), some additional work is needed on its scalability in practice, for example through sampling methods.

\subsection{Activation}
\label{m:activation}
The \emph{Activation} metric expresses whether the issued recommendations are aimed at inspiring the users to take action. A score close to 1 indicates a high amount of activating content, whereas a score close to 0 indicates more neutral content.
\subsubsection{Explanation.} 
The way in which an article is written may affect the reader in some way. An impartial article may foster understanding for different perspectives, whereas an emotional article may activate them to undertake action. A lot of work has been done on the effect of emotions and affect on the undertaking of collective group action. This holds especially for anger, in combination with a sense of group efficacy~\cite{van2004put}. But positive emotions play a role too; for example, "joy" elicits the urge to get involved, and "hope" to dream big~\cite{fredrickson2013positive}. The link between emotions, affect and activation is described well by \citet{papacharissi206affect}: \emph{"...for it is affect that provides the intensity with which we experience emotions like pain, joy, and love, and more important, the urgency to act upon those feelings"}. The Activation metric aims to capture this by measuring the strength of emotions expressed in an article.
\subsubsection{In the context of democratic recommenders.}
The Activation metric is relevant in three of the four different models. The Deliberative model aims for a common consensus and debate, and therefore would give a certain measure of prominence to impartial articles with low Activation scores. The Participatory model fosters the common good and understanding, and aims to facilitate users in fulfilling their roles as citizens, undertaking action when necessary. This leads to a slightly wider value range; some activating content is desirable, but nothing too extreme. The Critical model however focuses specifically on emotional and provocative content to challenge the status quo. Here very high values of Activation should be expected.
\subsubsection{Operationalization.} The Circumplex Model of Affect~\cite{russell2003core} describes a dimensional model where all types of emotions are expressed using the terms \emph{valence} and \emph{arousal}. \emph{Valence} indicates whether the emotion is positive or negative, while \emph{arousal} refers to the strength of the emotion and to what extent it expresses action. Following this, for example, 'excitement' has a positive valence and arousal, whereas 'bored' is negative for both. Based on the theory described above a number of "sentiment analysis" tools have been developed, which typically have the goal of identifying whether people have a positive or negative sentiment regarding a certain product or issue. For example, \citet{hutto2014vader} provides a lexicon-based tool that for each input piece of text outputs a compound score ranging from -1 (very negative) to 1 (very positive). The absolute values of these scores can be used as an approximation of the arousal and therefore be used to determine the Activation score of a single article. Then, the total Activation score of the recommender system should be calculated two-fold. The average Activation score of the items recommended to each user provides a baseline score for whether the articles overall tend to be activating or neutral. Next, the issued recommendations are compared to the available pool of data as follows:
\[Activation(p,q) = (|polarity(q)| - |polarity(p)|)/2\]
Here $p$ denotes the set of all available articles in the pool, and $q$ those in the recommendation. 
For both sets we take the mean of the absolute polarity value of each article, which we use as an approximation for Activation. We subtract the mean from the available pool of articles from the mean of the recommendation set, which maps to a range of $[-1,1]$. A value lower than zero indicates that the recommender system shows less activating content than was available in the pool of data, and therefore favors more neutral articles. Values higher than zero show the opposite; the recommendation sets contained proportionally more activating content than was available in the pool
.
\subsubsection{Limitations.} Of principle importance is the impact that the article's text has on the reader. However, as we have no direct way of measuring this, we hold to the assumption that a strongly emotional article will also cause similarly strong emotions in a reader, which again translates into higher willingness to act. It must also be noted that people may respond differently to different emotions (for example, anger may incite either approach (action) or avoidance (inaction) tendencies)~\cite{schuck2015news}. We therefore see this approach as an approximation of the concept of activation, affect and emotion in articles, until such a time when more research in the topic allows us to be more nuanced in our perceptions.

\subsection{Representation}
\label{m:representation}
The \emph{Representation} metric expresses whether the issued recommendations provide a balance of different opinions and perspectives, where one is not unduly more or less represented than others. A score close to zero indicates a balance, where the model of democracy that is chosen determines what this balance entails, whereas a higher score indicates larger discrepancies. 
\subsubsection{Explanation.} Representation is one of the more intuitive interpretations of diversity. Depending on which model of democracy is chosen, news recommendations should contain a plurality of different opinions. Here we care more about \emph{what} is being said than \emph{who} says it, which is the goal of the final metric, Alternative Voices. In order to define what it means to provide a balance of opinions, one needs to refer back to the different models and their goals.
\subsubsection{In the context of democratic recommenders.}
The Participatory model aims to be \emph{reflective} of "the real political world". Power relations that are therefore present in society should also be present in the news recommendations, with a larger share in the Representation for the more prevalent opinions. On the other hand, the Deliberative model aims to provide an \emph{equal} overview of all opinions without one being more prevalent than the other. One should keep in mind that not each recommendation by itself needs to provide an overview of the full spectrum, but rather that the aggregate of all recommendations should be complete and unbiased. The Critical model has a large focus on shifting power balances, and it does so by giving a platform to underrepresented opinions, thereby promoting an \emph{inverse} point of view. In doing this, the Critical model also strongly considers the characteristics of the opinion holder, specifically whether they are part of a minority group or not, though this is the goal of the last metric, Altenrative Voices.
\subsubsection{Operationalization.}
Representation, and Alternative Voices as well, rely strongly on the correct and complete identification of the opinions and opinion holders mentioned in the news. Though there is research available on the usage of Natural Language patterns to extract opinion data from an article's text~\cite{pareti2013automatically}, additional work is necessary on its applicability in this context. For example, it is of significant importance that not one type of opinion or opinion holder is systematically missed. Once the quality of the extraction is relatively certain, additional work is also necessary on the placement of opinions relative to each other; for example, which opinions are in favor, against or neutral on a statement, and how are these represented in the recommendations. This task is extremely complex, even for humans. In the meantime approximations can be used, for example by considering (spokespersons of) political parties and their position on the political spectrum. This can be done through manual annotations, with hardcoded lists of politicians and their parties, or automatically by for example querying Wikidata for information on persons identified through Named Entity Recognition. 
To calculate the Representation score, we once again use the Kullback-Leibler Divergence, but this time on the different opinion categories in the recommendations versus the available pool of data: 
\[ Representation_{(p,q)} = \sum_{o}^{} p(o) log \frac{p(o)}{\tilde{q}(o|u)} \]
This calculation is similar to the one in Section \ref{m:calibration}. However, $o$ indicates the different opinions in the data; $p(o)$ represents the proportion of the times this opinion was present in the overall pool of data, whereas $\tilde{q}(o|u)$ represents the proportion of times user $u$ has seen this opinion in their recommendations. A score of 0 means a perfect match between the two, which means that the opinions shown in the recommendations are perfectly representative of those in society. When following the Participatory's model \emph{reflective} point of view we want this value to be as close to zero as possible, as being representative of society is its main goal. However, when following one of the other models, we have to make some alterations on the distributions expressed by $p$. The Critical model's \emph{inverse} point of view aims for the recommendations to diverge as much from the power relations in society as possible. However, since very small differences in distributions can result in a very large KL divergence, simply maximizing the KL divergence is not sufficient. Instead, we inverse the distribution of opinions present in $p$. Similarly, when choosing the Deliberative model, we want all opinions in the recommendations to be equally represented, and therefore we choose $p$ as a uniform distribution of opinions. This way, for each of the different approaches holds that the closer the divergence is to zero, the better the recommendations reflect the desired representation of different opinions.  
For each of the reflective, inverse and equal approaches, the aggregated Representation score is obtained by averaging the Representation score over all recommendations issued to all users.
\subsubsection{Limitations.} Kullback-Leibler divergence treats each category as being independent, and does not account for opinions and standpoints that may be more or less similar to other categories. 

\subsection{Alternative Voices}
The \emph{Alternative Voices} metric measures the relative presence of people from a minority group. A higher score indicates a proportionally larger presence.
\subsubsection{Explanation.} Where Representation is largely focused on the explicit content of a perspective (the \emph{what}), Alternative Voices is more concerned with the person holding it (the \emph{who}), and specifically whether this person or organisation is one of a minority or an otherwise marginalised group that is more likely to be underrepresented in the mainstream media. What exactly entails a minority is rather vaguely defined. Article 1 from the 1992 United Nations Minorities Declaration refers to minorities “a non-dominant group of individuals who share certain national, ethnic, religious or linguistic characteristics which are different from those of the majority population", though there is no internationally agreed-upon definition. In practice, this interpretation is often extended with gender identity, disability and sexual orientation. 
A major challenge of the Alternative Voices metric lies in the actual identification of a minority voice. Though there are a number of studies that aim to detect certain characteristics of minorities from textual data, such as predicting a person's ethnicity and gender based on their first and last name~\cite{sood2018predicting}, there are no approaches that 1) model all minority characteristics or 2) perform well consistently. This process needs significant additional and most importantly multidisciplinary research, with a large focus on ensuring that doing this type of analysis does not lead to unintended side effects and unwanted consequences. For example, \citet{keyes2018misgendering} shows that current studies typically treat gender classification as a purely binary problem, thereby systematically leaving out and wrongly classifying transgender people. Similarly, \citet{hanna2020towards} argues that race and ethnicity are strongly social constructs that should not be treated as objective differences between groups. This topic, typically referred to as (algorithmic) Fairness, is an active research field that aims to counter bias and discrimination in data-driven computer systems. One thing is for certain: any recommender system that actively promotes one type of voice over another should make very explicit on what criteria and following which methods it does this. Following this both the identification and the way its algorithms use this information must be fully transparent and auditable. However, for the remainder of this section we will assume that we do have a proper way of identifying people from a minority group, either through manual annotation or automatic extraction.
\subsubsection{In the context of democratic recommenders.}
The Alternative Voices metric is naturally most significant in the Critical model, which aims to provide a platform to voices that would otherwise go unheard, and therefore has a large focus on the opinions and perspectives from minority groups. To a lesser extent, the same holds for the Participatory model, which aims to foster tolerance and empathy and encourages citizens to act.
\subsubsection{Operationalization.}
The discussion around Fairness in machine learning systems has lead, among others, to a number of definitions of the concept. For the operationalization of Alternative Voices we adapt Equation 10 of \citet{burke2018balanced} for our purposes:
\[ Alternative Voices = \frac{q^{+}/p^{+}}{q^{-}/p^{-}} \]
Here $q^{+}$ denotes the number of mentions of people belonging to a protected group in the recommendations, whereas $p^{+}$ denotes the number of mentions of people belonging to a protected group in all the available articles. $q^{-}$ and $p^{-}$ denote similar mentions, but for people belonging to the unprotected group.
Though the example given in \citet{burke2018balanced} describes the equation being used to identify whether loans from protected and unprotected regions appear equally often, it is also directly applicable to our notion of Alternative Voices; however, rather than counting regions being recommended, we count the number of times that people from minority (protected) versus majority (unprotected) groups are being mentioned in the news. This function maps to 1 when there is a complete balance between people from the protected and the unprotected groups; proportionally to the total number of times they appear in the data set, they appear a similar number of times in the recommendations. When the value is larger than 1 more people from unprotected groups appear in the recommendation set, whereas lower than 1 means they appear less. \\
Again, the aggregate score consists of the average Alternative Voices score over all recommendations issued to all users.
\subsubsection{Limitations.} A major caveat of this approach is that it assumes that the mere mentioning of minority people is enough to serve the goals of the Alternative Voices metric. This disregards the fact that these people may be mentioned but from another person's perspective, or in a negative light. Further research should focus on not only identifying a person from a minority group, but also whether they are mentioned as an active or passive agent.

\begin{table*}[ht]
\begin{tabular}{l|llllll}
\textbf{}                                      & \cellcolor[HTML]{EFEFEF}\textbf{\begin{tabular}[c]{@{}l@{}}Calibration\\ (topic)\end{tabular}} & \cellcolor[HTML]{EFEFEF}\textbf{\begin{tabular}[c]{@{}l@{}}Calibration\\ (style)\end{tabular}} & \cellcolor[HTML]{EFEFEF}\textbf{Fragmentation} & \cellcolor[HTML]{EFEFEF}\textbf{Affect} & \cellcolor[HTML]{EFEFEF}\textbf{Representation} & \cellcolor[HTML]{EFEFEF}\textbf{Alternative Voices} \\ \hline
\cellcolor[HTML]{EFEFEF}\textbf{Liberal}       & High                                                                                           & High                                                                                           & High                                           & -                                       & -                                               & -                                          \\ \hline
\cellcolor[HTML]{EFEFEF}\textbf{Participatory} & Low                                                                                            & High                                                                                           & Low                                            & Medium                                  & Reflective                                      & Medium                                     \\ \hline
\cellcolor[HTML]{EFEFEF}\textbf{Deliberative}  & -                                                                                              & -                                                                                               & Low                                            & Low                                     & Equal                                           & -                                          \\ \hline
\cellcolor[HTML]{EFEFEF}\textbf{Critical}      & -                                                                                              & -                                                                                               & -                                              & High                                    & Inverse                                               & High                                      
\end{tabular}
\caption{Overview of the different models and expected value ranges for each metric. Note that for the metrics reflecting distance of a distribution (Calibration and Representation), a "High" target value actually means that the resulting value should be close to zero.}
\label{tab:table_overview}
\vspace{-4mm}
\end{table*}

\section{General Limitations}
\label{sec:limitations}
Though all of the metrics described in Section \ref{sec:evaluation_metrics} already mention the limitations of that metric specifically, this section describes a number of the limitations of this method as a whole.
\subsection{Ordering}
Of the currently specified metrics, only Fragmentation takes the ordering of the items in the recommendation into account. However, the top result in a recommendation is of significantly more importance than the result in place 10. In future work, the other metrics should be extended in such a way that they reflect this.
\subsection{Engagament and Contestation}
One important aspect of the different models that is currently not reflected in the metrics is the amount of Engagement or more specifically Contestation they spark in their userbase. Ultimately, the aim of all models is to spark civic discussions and action, albeit to different ends (e.g. whereas in the Deliberative model the deliberation itself is a major aim, in the Critical model engagement and contestation must ultimately result in challenging established distributions of power and decision routines). This level of Engagement, which could for example be approached by counting the number of replies in a discussion thread or the number of shares on social media, can only be measured after the recommendations have been issued and are often already a standard part of the online evaluation of the performance of a recommender system~\cite{beel2013onlineoffline}. 
\subsection{Formalism Trap}
Many of the concepts described here are susceptible to the Formalism Trap described in \cite{selbst2019fairness}, which is defined as the \emph{"[f]ailure to account for the full meaning of social concepts [...], which can be procedural, contextual and contestable, and cannot be resolved through mathematical formalisms"}. Though our approach aims to model concepts founded in social science theories, they are merely approximations and to a large extent simplifications of very complex and nuanced subjects that have been contested and debated in the social sciences for decades. To claim our approach comes close to covering these subtleties would be presumptuous - however, we do believe it is necessary to provide a starting point in the modeling of concepts that have so far been neglected in the evaluation of news recommendations. The pitfalls of this trap should be mitigated by always providing full transparency on how these concepts are implemented, on what kind of data they are based, and most importantly on how they should (and should not) be interpreted.
\subsection{Bias in the dataset}
The metrics presented in Section \ref{sec:evaluation_metrics} typically rely on measuring a difference between the set of  recommended items and the full set of articles that were available, the reading history of the user in question or among users. What it does not do is account for inherent bias in the overall dataset, 
though the possibility of exposure diversity depends on the availability of content in the pool. If the quality and diversity of the pool is low, recommenders have insufficient options to provide good recommendations. That means exposure diversity ultimately is dependent on external diversity. Detecting such a bias in the dataset rather than in the produced recommendations and undertaking steps to remedy this needs additional work.

\subsection{Inherent limits to value by design approaches}
Finally, it is important to be mindful of another lesson from the general diversity by design literature, namely that there are also certain limits to value sensitive design, and for our case: the extent to which diversity as a normative concept can be operationalized in concrete recommender design.  This can have to do with the sheer difficulty of translating certain aspects of diversity, but also with the trade-offs between values that optimizing for exposure diversity can involve. Examples of this are commercial constraints and the need to optimize for profit rather than for diversity metrics, but also the limited effectiveness of recommenders in actually steering user choices. As \citet{friedmansurvey2017} explain: "a given technology is more suitable for certain activities and more readily supports certain values while rendering other activities and values more difficult to realize."

\section{Implementation}
\label{sec:implementation}
We are working on the implementation of the concepts and metrics discussed here in an open source tool\footnote{Link taken out during review}. The goal of this tool is to implement the metrics described in this paper as evaluation metrics for recommender design, and in doing so enable media companies to evaluate the performance of their own recommendations against those of several baseline recommendations.
\subsection{Approach}
\begin{figure}[bp]
  \includegraphics[width=\linewidth]{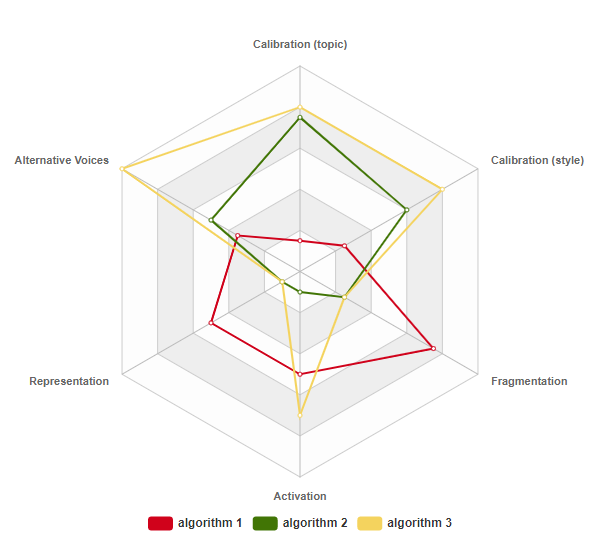}
  \caption{Example visualization of the proposed metrics within the tool.}
  \label{fig:spider}
\end{figure}

By making comparisons between the different recommender approaches, media companies should be able to draw conclusions about which recommender strategy fits their mission best. Figure \ref{fig:spider} shows an example of what such a comparison could look like. Here, three different algorithms are compared. Algorithm 1 is most suitable for recommender systems following the Liberal model, as it is well-calibrated to the user's needs. Algorithm 2 is more suited for the Deliberative model (low Activation and non-diverging Representation) and algorithm 3 for the Critical model (high Activation and Alternative Voices). In practice these differences are unlikely to be this distinctive; by also comparing the performance of these algorithms to very simple recommendation approaches, such as a random recommender, the media company can also draw conclusions about where the recommender simply reflects the available data, and where it significantly influences the type of data that is shown. By making these visualizations as intuitive as possible, they should facilitate the discussion between data science teams, editors and upper management around this topic. To make this approach reusable and broadly applicable, it should be implemented and tested on both a benchmark set such as~\cite{wu2020mind} and in a real-life setting. 
We are in contact with multiple media companies, to inform them about the different models of democracy, facilitate the discussion around this subject, and stimulate the implementation of our tool. Simultaneously this topic is continuously being discussed with experts from many different disciplines, as happened for example during a Dagstuhl Workshop\footnote{link taken out during review}.

\subsection{Guidelines for adoption}
\label{sec:guidelines}
The ultimate goal of this paper is to propose notions that could be incorporated in recommender system design. In our vision, media companies could approach this in the following steps:
\begin{enumerate}
    \item Determine which model of democracy to follow \\
    Following the different models described in Section \ref{sec:background}, the media company in question should decide which model of democracy the recommender system should reflect. This is something that should be decided in active discussion with the editorial team, and directly in line with the media company's mission.
    \item Identify the corresponding metrics \\
    Use Table \ref{tab:table_overview} to determine which metrics are relevant, and what the expected value range for each metric is. For example, when choosing to follow the Deliberative model, the recommender system should optimize for a low Fragmentation, low Activation and equal Representation. Similarly, for the Critical model, it should optimize for high Activation, inverse Representation and high Alternative Voices.
    \item Implement into recommender design \\
    Here it is of key importance to determine the relative importance of each metric, and how to make a trade-off between recommender accuracy and normative diversity. For example, \citet{mehrotra2018towards} details a number of approaches to combining Relevance and Fairness in Spotify's music recommendation algorithm, and this approach can also be applied in the trade-off between accuracy and the metrics relevant for the chosen model.   
\end{enumerate}
We do not consider these metrics to be the final "truth" in the identification of diversity in news recommendations. For example, it is possible and acceptable that during discussions, a media company decides that they aim to follow the Deliberative model of democracy. However, rather than focusing on the equal Representation of opinions in the recommendations, they decide to focus on providing a good balance of background- and opinion pieces, either for the sake of practicality or because this is more in line with their mission. The metrics and their operationalizations should serve as inspiration and a starting point for discussion, not as restrictions or set requirements for "good" recommender design.

\section{Discussion}
In this paper we have translated normative notions of diversity into five metrics. Each of the metrics proposed here is relevant in the context of democratic news recommenders, and combined they form a picture that aims to be expressive of the nuances in the different models. However there is still a lot of work to be done, both in terms of technical feasibility and in undertaking steps to make diversity of central importance for recommender system development.

At the basis of our work is that we believe diversity is not a single absolute, but rather an aggregate value with many aspects. In fact, we argue that what constitutes 'good' diversity in a recommender system is largely dependent on its goal, which type of content it aims to promote, and which model of the normative framework of democracy it aims to follow. As none of these models is inherently better or worse than the others, we believe that a media company should take a normative stance and evaluate their recommender systems accordingly. 

Different fields and disciplines may have very different notions of the same concept, and navigating these differences is a process of constant negotiation and compromise, but also of expectation management. Abstract concepts such as diversity may never be fully captured by the cold, hard numbers that recommender system practitioners are used to. As technological advances take on an ever more central role in society, the necessity to bridge this gap and make such concepts more concrete also arises. Social sciences, humanities and computer science will need to meet in the middle between abstract and concrete, and work together to create ethical and interpretable technologies. This work is not a final conclusion on how diversity can be measured in news recommendations, but rather a first step in forming the bridge between the normative notion of diversity and its practical implementation.
\label{sec:conclusions}

\bibliographystyle{ACM-Reference-Format}
\bibliography{paper}

\end{document}